# Background Independent Quantum Mechanics, Classical Geometric Forms and Geometric Quantum Mechanics-I

Aalok*

Department of Physics, University of Rajasthan, Jaipur 302004, India;
and Jaipur Engineering College and Research Centre (JECRC), Jaipur 303905, India.

**Abstract**: The geometry of the symplectic structures and Fubini-Study metric is discussed. Discussion in the paper addresses geometry of Quantum Mechanics in the classical phase space. Also, geometry of Quantum Mechanics in the projective Hilbert space has been discussed for the chosen Quantum states. Since the theory of classical gravity is basically geometric in nature and Quantum Mechanics is in no way devoid of geometry, the explorations pertaining to more and more geometry in Quantum Mechanics could prove to be valuable for larger objectives such as understanding of gravity.

*E-mail*: aalok@uniraj.ernet.in

# 1. Introduction

The discussion in this paper investigates the possible geometric consequences of the background independent Quantum Mechanics and an extended framework of Quantum Mechanics as described in a comprehensive perspective in a recent paper [1].

As a quantum system evolves in time the state vector changes and it traces out a curve in the Hilbert space *H*. Geometrically, the evolution is represented as a closed curve in the projective Hilbert space *P* [1-8, 15]. Researchers studying gravity have also shown considerable interest in the geometric structures in quantum mechanics in general and projective Hilbert space in specific [1, 6, 9-14]. In the light of recent studies [1, 2, 9-14] of geometry of the quantum state space, the need and call for further extension of standard geometric quantum mechanics is irresistible. Thus an intensive follow up will be academically rewarding [1]. Classical mechanics has deep roots in (symplectic) geometry while quantum mechanics is essentially algebraic. However, one can recast quantum mechanics in a geometric language, which brings out the similarities and differences between two theories [6]. The idea is to pass from the Hilbert space to the space of rays, which is the "true" space of states of quantum mechanics. The space of rays- or the projective Hilbert space is in particular, a symplectic manifold, which is equipped with a *Kähler* structure. Regarding it as a symplectic manifold, one can recast the familiar constructions of classical mechanics.

The discussion in the present paper addresses various aspects of geometry of Quantum Mechanics. The discussion in the section 2 of the paper addresses geometry of Quantum Mechanics in the classical phase space. And the discussion in the section 3 of the present paper addresses geometry of Quantum Mechanics in the projective Hilbert space.

At the out-set, we briefly review the background and the settings in which various exercises have been carried out in this paper.

The *distance* on the projective Hilbert space is defined in terms of metric, called the metric of the ray space or the projective Hilbert space $\mathcal{P}$, is given by the following expression in Dirac's notation:

$$ds^2 = 4\left(1 - |\langle \Psi_1 | \Psi_2 \rangle|^2\right) \equiv 4\left(\langle d\Psi | d\Psi \rangle - \langle d\Psi | \Psi \rangle \langle \Psi | d\Psi \rangle\right). \tag{1}$$

This can be regarded as an alternative definition of the Fubini-Study metric, valid for an infinite dimensional $\mathcal{H}$.

The metric in the ray space is now being referred by physicists as the background independent and space-time independent structure, which can play an important role in the construction of a potential "theory of quantum gravity". The demand of background independence in a quantum theory of gravity calls for an extension of standard geometric quantum mechanics [1, 9-14]. The metric structure in the projective Hilbert space is treated as background independent and space-time independent geometric structure. It is an important insight which can be a springboard for our proposed background independent generalization of standard quantum mechanics. For a generalized coherent state, the FS metric reduces to the metric on the corresponding group manifold [11-13]. Thus, in the wake of ongoing work in the field of quantum geometric formulation, the work in the present paper may prove to be very useful. The probabilistic (statistical) interpretation of QM is thus inherent in the metric properties of $\mathcal{P}(\mathcal{H})$ [3-5, 9-14]. The unitary time evolution is related to the metrical structure [3] with Schrödinger's equation in the guise of a geodesic equation on $CP(N)$. Thus the metric of quantum state space is found to be independent of choice of quantum evolution, relativistic or non-relativistic.

The metric of quantum state space has been identified as background independent (BI) metric structure [1, 2, 9-14]. However, by appearance itself the invariance of the geometric structure in equation (1) is apparent, irrespective of the choice of state function.

In the context of complex projective space $CP(N)$, due to $Diff(\infty, C)$ symmetry, the "coordinates" $Z^a$ while representing quantum states, make no sense physically, only quantum events do, which is the quantum counterpart of the corresponding statement on the meaning of space-time events in General Relativity (GR). Probability is generalized and given by the notation of diffeomorphism invariant distance in the space of quantum configurations. The dynamical equation is a geodesic equation in this space. Time, the evolution parameter in the generalized Schrödinger equation, is yet not global and is given in terms of the invariant distance. The basic point as threshold of the background independent quantum mechanics (BIQM) is to notice that the evolution equation (the generalized Schrödinger equation) as a geodesic equation, can be derived from an Einstein-like equation with the energy-momentum tensor determined by the holonomic non-abelian field strength $F_{ab}$ of the $Diff(\infty-1, C) \times Diff(1, C)$ type and the interpretation of the Hamiltonian as a charge.

Such an extrapolation is logical since $CP(N)$ is an Einstein space, and its metric obeys Einstein's equation with a positive cosmological constant given by:

$$R_{ab} - \frac{1}{2} R g_{ab} - \Lambda g_{ab} = 0. \qquad (2)$$

The diffeomorphism invariance of the new phase space suggests the following dynamical scheme for the (BIQM) as:

$$R_{ab} - \frac{1}{2} R g_{ab} - \Lambda g_{ab} = T_{ab}, \tag{3}$$

with $T_{ab}$ be given as above.

Furthermore, $\nabla_a F^{ab} = \frac{1}{2\Delta E} H u^b$. (4)

The last two equations imply *via* the Bianchi identity, a conserved energy-momentum tensor: $\nabla_a T^{ab} = 0$. (5)

This taken together with the conserved "current" as:

$$j^b = \frac{1}{2\Delta E} H u^b, \text{ and } \nabla_a j^a = 0; \tag{6}$$

implies the generalized geodesic Schrödinger equation. Thus equation (3) and (4), being a closed system of equations for the metric and symplectic structure do not depend on the Hamiltonian, which is the case in ordinary quantum mechanics. By imposing the conditions of homogeneity and isotropy on the metric by means of number of Killing vectors, the usual quantum mechanics can be recovered [6, 11-13]. And this limit does not affect the geodesic equation:

$$\frac{du^a}{ds} + \Gamma^a_{bc} u^b u^c = \frac{1}{2\Delta E} Tr(H F^a_b) u^b, \tag{7}$$

due to the relation $\hbar d\tau = 2\Delta E dt$. (8)

The reformulation of the geometric QM in this background independent setting gives us lot of new insights. The utility of the BIQM formalism is that gravity embeds into quantum mechanics with the requirement that the kinematical structure must remain compatible with the generalized dynamical structure under deformation. The requirement of diffeomorphism invariance places stringent constraints on the quantum geometry. We must have a strictly almost complex structure on the generalized space of quantum events.

The symmetries as described by the quotient set $CP(N) = \frac{U(N+1)}{U(N) \times U(1)}$, have limitations.

Thus the only alternative that seems to satisfy the almost complex structure is the Grassmannian. By the correspondence principle, the generalized quantum geometry must locally recover the canonical quantum theory encapsulated in $\mathcal{P}(\mathcal{N})$ and also allows for mutually compatible metric and symplectic structure, supplies the framework for the dynamical extension of the canonical quantum theory.

The Grassmannian: $Gr(C^{n+1}) = Diff(C^{n+1})/Diff(C^{n+1}, C^n \times \{0\})$. (9)

This space is generalization of $\mathcal{P}(\mathcal{N})$. The Grassmannian is a gauged version of complex projective space, which is the geometric realization of quantum mechanics. The utility of this formalism is that gravity embeds into quantum mechanics with the requirement that the kinematical structure must remain compatible with the generalization dynamical structure under deformation. The quantum symplectic and metric structure, and therefore the almost complex structure, are themselves fully dynamical.

It has been underlined [11-13] time and again, that if the induced classical configuration space is to be the actual space of space-time, only a special quantum system will do. We are thus induced to make the state manifold suitably flexible by doing general relativity on it. The resultant metric on the Hilbert space is generally curved with its distance function modified, an extended Born' rule and hence a new extended Quantum Mechanics [11-13].

Geometry of the Quantum Mechanics with its relevance in the studies of gravity has been explored [7, 8, 16] earlier too. Holonomy in quantum evolution was described by Simon

Berry [8]. Moreover, the holonomies of geometrical structures on the projective space and associated Cartan's geometric forms on the *Kähler* manifold were explored by Don Page [16]. The present paper too aims to unfold various geometric aspects of Quantum Mechanics that have yet not been explored so far.

## 2. The classical geometry *vis-à-vis* geometric quantum mechanics

In the following discussion, the geometry of Quantum Mechanics is described in the classical phase space. In some sense, the present work is an extension of the explorations in the reference [4].

We discuss here the classical limit of the symplectic structures and Fubini-Study metric on $\mathcal{P}$ by considering states which have the same uncertainties $\Delta q$, and $\Delta p$ in the position and momentum and product $(\Delta q)(\Delta p)$ has the minimum value $\hbar/2$. Such states are of the form:

$$\Psi_{\vec{q},\vec{p}}(\vec{x}) = [2\pi(\Delta q)^2]^{-\frac{1}{4}} \exp\left(-\frac{(\vec{x}-\vec{q})^2}{(2\Delta q)^2} + \frac{i}{\hbar}\vec{p}.\vec{x}\right). \tag{10}$$

Where, the only arbitrariness is in the choice of the average position $\vec{q}$, average momentum $\vec{p}$, the width $\Delta q$ and the phase of the normalization constant, chosen to be zero, of this Gaussian wave packet. In the classical limit, the quantum system has a large enough mass for the spreading of the wave packet to be negligible during a time interval of interest. Then its state may be assumed to remain as a Gaussian wave packet with $(\vec{q},\vec{p})$ changing with time, to a very good approximation.

Physically, it is possible in principle to move a quantum system along an arbitrary curve in $\mathcal{C}$ by making a dense sequence of measurements [4]. Therefore, the sub-manifold $\mathcal{C}$ of $\mathcal{P}$ consisting of the states (10), for a fixed $\Delta q$, can be identified with the classical phase space $\{(\vec{q}, \vec{p})\}$. Moreover, due to the spread of the wave packet, the tangent space at each point of $C$ may be *locally* identified with the classical phase space.

Consider a cyclic motion $(\vec{q}(\lambda), \vec{p}(\lambda))$, that is a closed curve **R** in $C$. If this motion is due to an evolution in the classical limit then $\lambda$ may be identified with time $t$. Then-

$$i\left\langle \tilde{\Psi}_{\vec{q}(\lambda),\vec{p}(\lambda)} \left| \frac{d\tilde{\Psi}_{\vec{q}(\lambda),\vec{p}(\lambda)}}{d\lambda} \right.\right\rangle = -\frac{1}{\hbar}\vec{q}\cdot\frac{d\vec{p}}{d\lambda}. \tag{11}$$

Using definition of the geometric phase prescribed as early as in the ref. [4],

$$\beta = i\oint_R \langle \tilde{\Psi} | d\tilde{\Psi} \rangle; \tag{12}$$

and Stoke's theorem, the geometric phase for this motion is found [3, 4] to be:

$$\beta = \frac{1}{\hbar}\int_S d\vec{p}\wedge d\vec{q}; \tag{13}$$

where $S$ is a surface in $C$ spanned by **R**.

And $\frac{1}{\hbar}d\vec{p}\wedge d\vec{q}$ is the restriction to $C$ of the symplectic form in $\mathcal{P}$.

Hence $\hbar\beta$ may be regarded, in this classical limit, as an oriented area in the classical phase- space with respect to its natural symplectic form and is therefore a canonical invariant [4].

Now, we consider a cyclic motion $(\vec{q}(s), \vec{p}(s))$ that is a closed curve **R** in $C$. If this motion is due to an evolution in the classical limit then, we can evaluate the corresponding Fubini-Study metric restricted to $C$. We choose an arbitrary curve $(\vec{q}(s), \vec{p}(s))$ in $\mathscr{C}$ of which the corresponding metric structure is defined as follow [4]:

$$ds^2 = \frac{1}{\Delta q^2}d\vec{q}^{\,2} + \frac{1}{\Delta p^2}d\vec{p}^{\,2}. \tag{14}$$

Where $(\Delta q)(\Delta p) \approx \hbar$ . (15)

Equation (14) is defined on the tangent space at each point of $C$ and is therefore not affected by the spread of the wave packet. The metric in the equation (14) on $C$ is flat, unlike the Fubini-Study metric in $\mathcal{P}$, which has constant curvature. This result was interesting because this was shown for the first time [4] that a metric in the classical phase space can be deduced from the fundamental principles of Quantum Mechanics.

This was not surprising that in selecting the sub-manifold $C$ of $\mathcal{P}$, Euclidean metric in the classical configuration space $X$ was used. But, we can treat the state vectors as abstract states in $\mathcal{H}$ and can think of a measuring instrument which measures position and momentum with the least uncertainty select the sub-manifold $C$, without explicitly invoking this metric. Moreover, for the classical Euclidean metrics in $X$ and $\mathcal{M}$, we may write

$C = X \times \mathcal{M}$, where $\mathcal{M}$ is the classical momentum space.

The usefulness of the metric in the equation (14), compared to the Euclidean metric in $X$, can be shown by considering the motion of a Gaussian wave packet in the classical limit of large mass so that its evolution may be regarded as a curve $C$ in $\mathcal{C}$. Suppose its projection $\gamma$ in $X$ has end points $\vec{q}_1, \vec{q}_2 \in X$. We are still interested to know- for what curve $C$, with well defined $\vec{q}_1$, $\vec{q}_2$, the distance along $C$ is determined by this metric is minimum. Obviously, it should be any curve for which the momentum $\vec{p}$ is constant, and $\gamma$ is a straight line.

But, if momentum $\vec{p}$ is not constant, the geometry of the following equation is nontrivial. We extrapolate the equation (14) and express it as:

$$\frac{\left(\frac{d\vec{q}}{ds}\right)^2}{(\Delta q)^2} + \frac{\left(\frac{d\vec{p}}{ds}\right)^2}{(\Delta p)^2} = 1, \tag{16}$$

with $(\Delta q)(\Delta p) \approx \hbar$.

Equation (16) represents elliptical geometry, if $\frac{d\vec{q}}{ds} = (\Delta q)\cos\theta$, and $\frac{d\vec{p}}{ds} = (\Delta p)\sin\theta$.

We now calculate the area of such an ellipse as:

$$A = \pi ab = \pi(\Delta p)(\Delta q),$$
$$\approx \pi\hbar. \tag{17}$$

This is precisely an oriented area in the classical phase space as discussed earlier in the reference [4].

However, it is a very complicated scenario. The equation $(\Delta q)(\Delta p) \approx \text{constant}$, follows geometry of a rectangular hyperbola. Moreover, the phase factor $\theta$ in parametric equation pair does not remain consistent. Here, in addition to the phase factor $\theta$, a geometric phase (say $\beta$) also arises during the quantum evolution as described in the equations (12) and (13). When we accommodate this emergence of additional phase factor in the parametric equation pair of ellipse, it causes precession in the elliptical geometry. Also, the presence of square of the derivatives as $\left(\frac{d\vec{q}}{ds}\right)^2$ and $\left(\frac{d\vec{p}}{ds}\right)^2$ in equation (16) is the signature of nonlinear geometry.

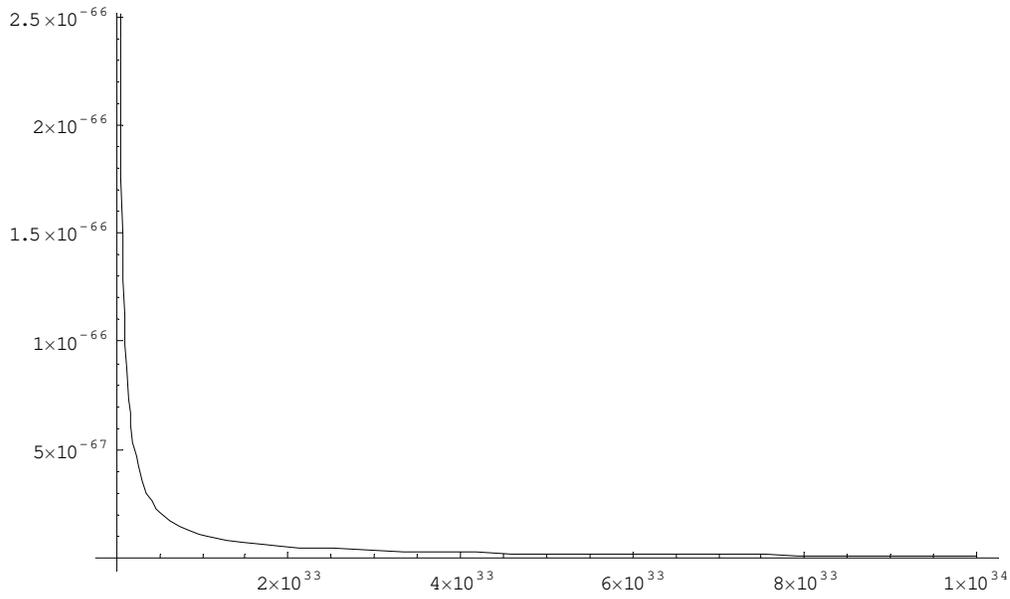

**Fig. 1(a)**. Graphical representation of geometry of the uncertainty relation $(\Delta q)(\Delta p) \approx \hbar$: a rectangular hyperbola.

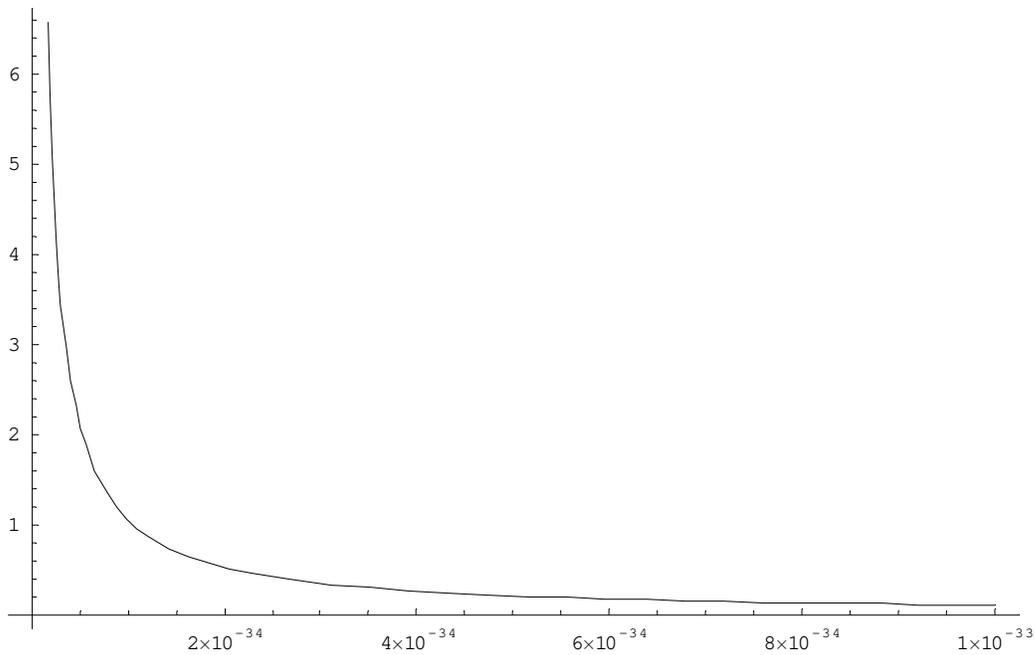

**Fig. 1(b).** Graphical representation of geometry of the uncertainty relation $(\Delta q)(\Delta p) \approx \hbar$: a rectangular hyperbola (with one of the parameters at Planck's scale).

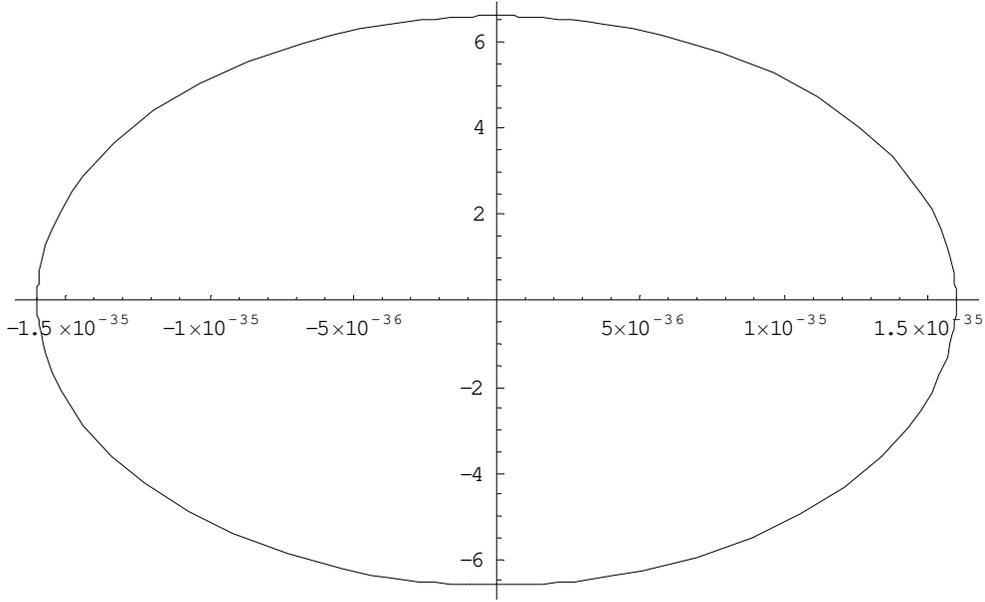

**Fig. 2**. Graphical representation of the elliptical geometry by using parametric forms for equation (16), with the constraint $(\Delta q)(\Delta p) \approx \hbar$: a computer simulation at Planck's scale.

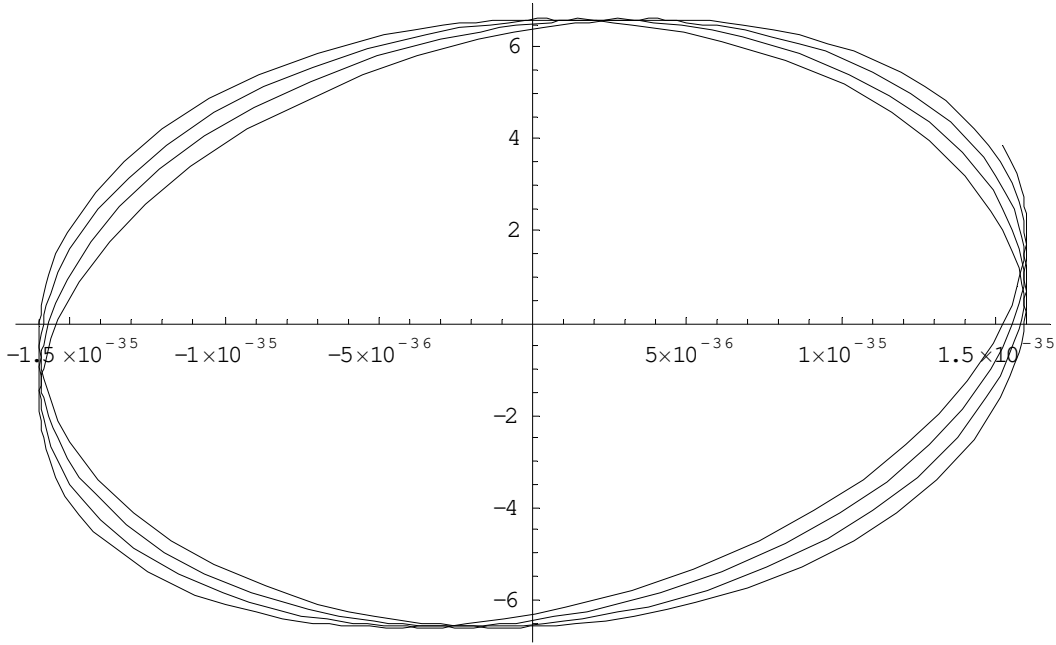

**Fig. 3**. The possible precession in the elliptical locus due to the constraint in the form of inequality $(\Delta q)(\Delta p) \geq \hbar$, and typical geometrical phase which arises in a quantum evolution: a computer simulation.

We know that kinematics and dynamics on an ellipse is rarely confined to a unique closed curve. Particularly, in the present case, where constraint relation is an inequality $(\Delta q)(\Delta p) \geq \hbar$, the precession in the elliptical locus is inevitable (see **Fig. 3**).

## 3. The geodesic equation and the length of a curve

The following discussion aims to explore the geometry of Quantum Mechanics in the projective Hilbert space. A comparison is also drawn with the classical geometric forms. In the studies of the projective Hilbert space, the *distance* [3-5, 17] and the *length* [17] have been defined. The *distance* as defined in equation (1) in terms of metric, is called the metric of the ray space or the projective Hilbert space $\mathcal{P}$. As discussed earlier, this is also known as Fubini-Study metric of the Ray space. Here, the invariant $ds$ can be regarded as the distance between points $p$ and $p'$ in the projective Hilbert space $\mathcal{P}$. And $|\Psi_1\rangle$, and $|\Psi_2\rangle$, are two normalised states contained in $p$ and $p'$; clearly with the condition $s(p, p') \geq 0$. This metric when restricted to $\mathcal{P}_1(\mathbb{C})$, is to be regarded as a 2-sphere with radius embedded in a real three-dimensional Euclidean space, and $s(p, p')$ is then the straight line, or better called geodesic distance between $p$ and $p'$ on this sphere [4, 5]. To be precise, the metric of quantum state space is a metric on the underlying manifold which the quantum states form or belong to, and therefore, it is different from the metric of space-time or any other metric associated with the quantum states. Also, it is to be noticed that this metric formalism is for pure states only. However, the metric of quantum states had been described for mixed states also [5]. The next step in this direction should be investigation of geodesic evolution. We begin with the geodesic equation for a quantum state evolving in time and given by variational calculus [15] as:

$$\frac{d^2\Psi}{ds^2} + u^2\Psi = 0. \tag{18}$$

Where $\langle\Psi|\Psi\rangle = 1$, $u^2 = \langle\dot{\Psi}|\dot{\Psi}\rangle$, $\tag{19}$

with $u$ being the speed of the system point in $\mathcal{P}$, and $\dot{\Psi}$ is the time derivative of the wave function. The geodesic in projective Hilbert space $\mathcal{P}$ can be defined as the one for which *length* of the path traced by the parallel-transported vector is minimum. The *length* of the curve traced by the parallel-transported state is given as [17]:

$$l = \int \left\langle \frac{\partial\Psi}{\partial t} \bigg| \frac{\partial\Psi}{\partial t} \right\rangle^{\frac{1}{2}} dt, \tag{20}$$

appropriately called the *horizontal lift* of the curve [15] corresponding to the quantum state under consideration. We discuss the nature of this *horizontal lift* corresponding to Hydrogen like wave functions.

Before we investigate the geometry on the quantum state space, we briefly discuss the length of an elliptical curve or perimeter of an ellipse and its variation with its eccentricity, it follows as:

$$l = 2\pi r \left[ 1 - \left(\frac{1}{2}\right)^2 \varepsilon^2 - \left(\frac{1\cdot 3}{2\cdot 4}\right)^2 \frac{\varepsilon^4}{3} - \left(\frac{1.3.5}{2.4.6}\right) \frac{\varepsilon^6}{5} \cdots \right], \tag{21}$$

or $l = 2\pi r f(\varepsilon)$; where $f(\varepsilon) = \left[ 1 - \left(\frac{1}{2}\right)^2 \varepsilon^2 - \left(\frac{1\cdot 3}{2\cdot 4}\right)^2 \frac{\varepsilon^4}{3} - \left(\frac{1.3.5}{2.4.6}\right) \frac{\varepsilon^6}{5} \cdots \right]$; $\tag{22}$

and quantity $2\pi f(\varepsilon)$ which is same for all elliptical curves, represents the horizontal lift of an ellipse. If we plot the behaviour of the horizontal lift of the elliptical curve with respect to variation in eccentricity ($\varepsilon$), we observe following pattern:

| $\varepsilon$ | $f(\varepsilon)$ |
|---|---|
| 0.1 | 0.997495 |
| 0.2 | 0.989995 |
| 0.3 | 0.977120 |
| 0.4 | 0.9588 |
| 0.5 | 0.934570 |
| 0.6 | 0.903925 |
| 0.7 | 0.866432 |
| 0.8 | 0.8208 |
| 0.9 | 0.766743 |
| 1 | 0.683593 |

**Table 1**. Eccentricity $\varepsilon$ for any conic and the corresponding values of $f(\varepsilon)$.

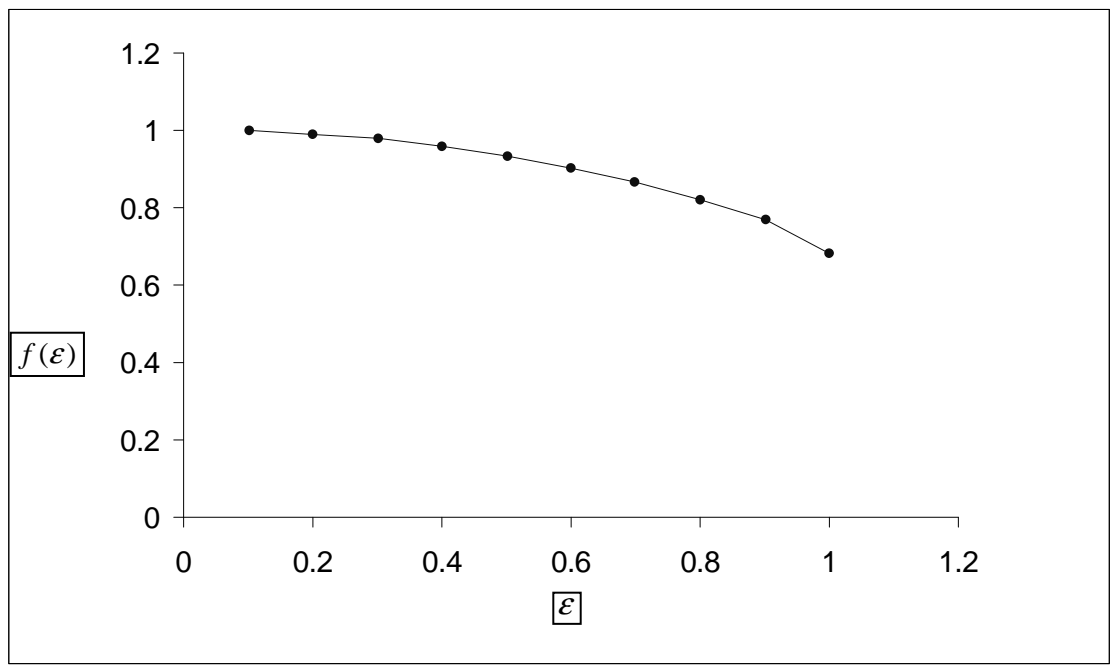

**Fig. 4**. Variation in $f(\varepsilon)$ with the variation in eccentricity $\varepsilon$.

Now, we examine the *length* or the *horizontal lift* of a curve with the quantum mechanical prescription:

$$l = \int \left\langle \frac{\partial \Psi}{\partial t} \middle| \frac{\partial \Psi}{\partial t} \right\rangle^{\frac{1}{2}} dt.$$

Thus the *horizontal lift* corresponding to the Hydrogen like wave function

$$\Psi_{200} = \left(\frac{1}{4\sqrt{2\pi}}\right)\left(\frac{1}{a_0}\right)^{\frac{3}{2}}\left(2 - \frac{r}{a_0}\right)e^{-\frac{r}{2a_0}}e^{-i\omega t}, \text{ where } a_0 \text{ is the Bohr radii,} \qquad (23)$$

could be evaluated as

$$l = \int_0^T \left\langle \frac{\partial \Psi_{200}}{\partial t} \middle| \frac{\partial \Psi_{200}}{\partial t} \right\rangle^{\frac{1}{2}} dt; \ 0 < r < R; \ 0 < \theta < \pi; \ 0 < \varphi < 2\pi; \ 0 < t < T. \qquad (24)$$

We identify this *horizontal lift* of the curve in terms of $f(\varepsilon)$ as follow:

$$l = 2\pi\left[1 - e^{-\frac{R}{a_0}}\left\{1 + \frac{R}{a_0} + \frac{R^2}{2a_0^2} + \frac{R^4}{8a_0^4}\right\}\right]^{\frac{1}{2}} = 2\pi f(\varepsilon). \qquad (25)$$

However, we do not know precisely the relation of eccentricity in terms of parameters $R$ and $a_0$, upon which the term $l = 2\pi f(\varepsilon)$ depends. But, for different values of $R$, the behaviour of function $\left[1 - e^{-\frac{R}{a_0}}\left\{1 + \frac{R}{a_0} + \frac{R^2}{2a_0^2} + \frac{R^4}{8a_0^4}\right\}\right]^{\frac{1}{2}}$ with respect to $\frac{a_0}{R}$ is same as the behaviour of $f(\varepsilon)$ for elliptical curves particularly when $\frac{a_0}{R} \ll 1$. Because $\frac{a_0}{R} \gg 1$ or $a_0 \gg R$ indicates behaviour of an open curve which never implies $f(\varepsilon) = 1$ and $l = 2\pi$.

If we plot the variation in $f(\varepsilon)$ graphically, it appears as follows:

| $\frac{a_0}{R}$ | $f(\varepsilon)$ |
|---|---|
| 0.04 | 1 |
| 0.05 | 0.9999 |
| 0.066 | 0.999 |
| 0.0833 | 0.991742 |
| 0.1 | 0.9697 |
| 0.111 | 0.944752 |
| 0.1428 | 0.8347 |
| 0.1666 | 0.73244 |

**Table 2**. The values of $\frac{a_0}{R}$ and the corresponding values of $f(\varepsilon)$, for the Hydrogen like wave function $\Psi_{200}$.

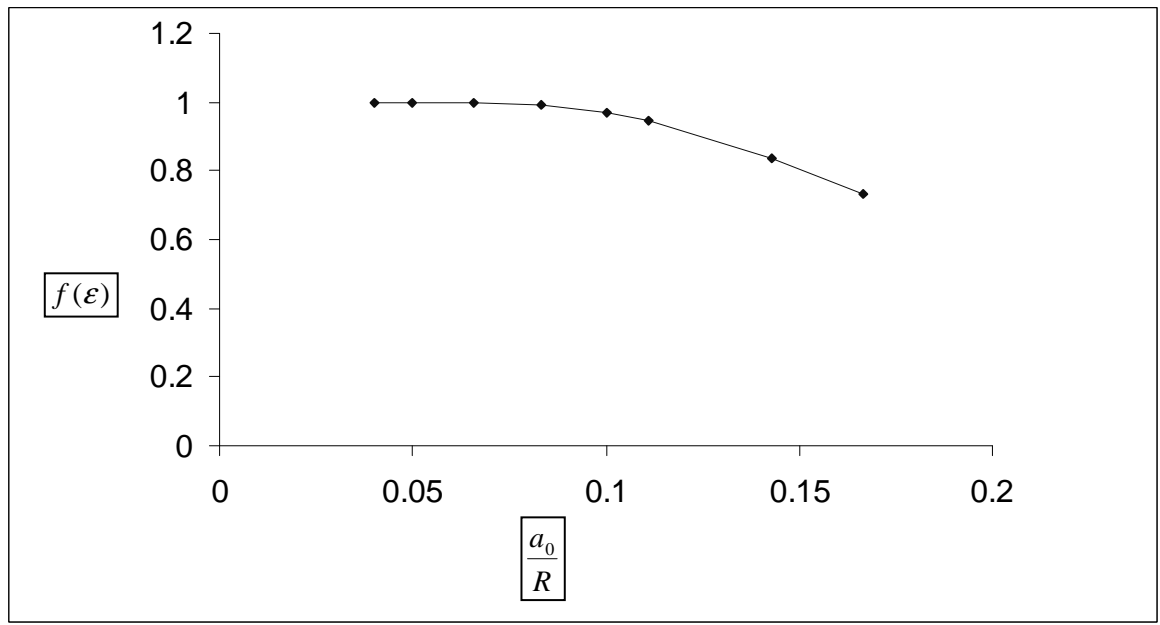

**Fig. 5**. Variation in $f(\varepsilon)$ with respect to $\frac{a_0}{R}$, corresponding to the wave function $\Psi_{200}$.

Here, each point on this curve denotes a possible closed (elliptical) curve trajectory. And the function $\left[1-e^{-\frac{R}{a_0}}\left\{1+\frac{R}{a_0}+\frac{R^2}{2a_0^2}+\frac{R^4}{8a_0^4}\right\}\right]^{\frac{1}{2}} = f(\varepsilon)$ attains a maximum value $f(\varepsilon)=1$, at $R = 24a_0$, and remains unity for all $R > 24a_0$. Also, we notice that the variation in the *horizontal lift* $l = 2\pi f(\varepsilon)$, are reminiscent of elliptical curves for values $R < 24a_0$, and the horizontal lift is maximal and is $2\pi$ for $R \geq 24a_0$, and the corresponding curves are all circular in nature.

The horizontal lift corresponding to the other Hydrogen like wave functions:

$$\Psi_{210} = \left(\frac{1}{4\sqrt{2\pi}}\right)\left(\frac{1}{a_0}\right)^{\frac{5}{2}} r e^{-\frac{r}{2a_0}} \cos\theta\, e^{-i\omega t},$$

$$\Psi_{211} = \left(\frac{1}{4\sqrt{2\pi}}\right)\left(\frac{1}{a_0}\right)^{\frac{5}{2}} r e^{-\frac{r}{2a_0}} \left(\frac{e^{i\varphi}}{\sqrt{2}}\right)\sin\theta\, e^{-i\omega t}, \text{and}$$

$$\Psi_{21-1} = \left(\frac{1}{4\sqrt{2\pi}}\right)\left(\frac{1}{a_0}\right)^{\frac{5}{2}} r e^{-\frac{r}{2a_0}} \left(\frac{e^{-i\varphi}}{\sqrt{2}}\right)\sin\theta\, e^{-i\omega t}; \tag{26}$$

is found to be

$$l = 2\pi\left[1-e^{-\frac{R}{a_0}}\left\{1+\frac{R}{a_0}+\frac{R^2}{2a_0^2}+\frac{R^3}{6a_0^3}+\frac{R^4}{24a_0^4}\right\}\right]^{\frac{1}{2}} = 2\pi f(\varepsilon). \tag{27}$$

The variation in the behaviour of the function $f(\varepsilon)$ is observed as follows:

| $\dfrac{a_0}{R}$ | $f(\varepsilon)$ |
|---|---|
| 0.04 | 1 |
| 0.05 | 0.9999 |
| 0.08 | 0.99732 |
| 0.1 | 0.985265 |
| 0.1111 | 0.97213 |
| 0.12 | 0.958 |
| 0.1428 | 0.9094 |
| 0.1666 | 0.8455 |
| 0.18 | 0.806891 |
| 0.2 | 0.748002 |

**Table 3**. Values of $\dfrac{a_0}{R}$ and the corresponding values of $f(\varepsilon)$, for Hydrogen like wave-functions $\Psi_{210}, \Psi_{211}, \Psi_{21-1}$.

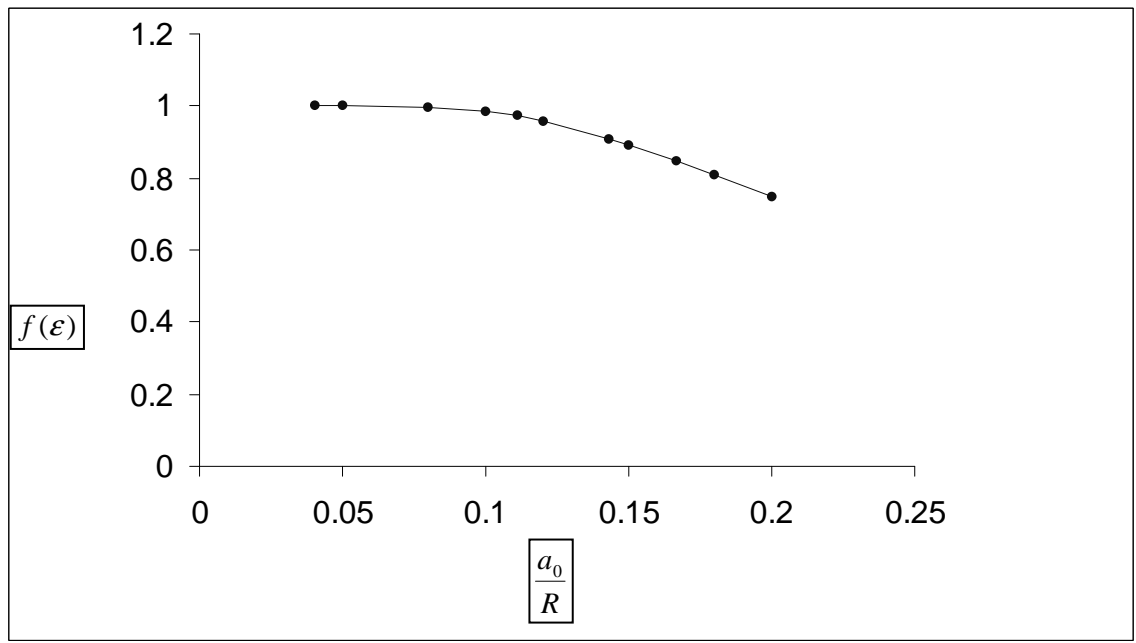

**Fig. 6**. Variation in $f(\varepsilon)$ with respect to $\dfrac{a_0}{R}$, corresponding to the wave functions $\Psi_{210}, \Psi_{211}, \Psi_{21-1}$.

Here, each point on this curve denotes a possible closed (elliptical) curved trajectory.

And the function $\left[1-e^{-\frac{R}{a_0}}\left\{1+\frac{R}{a_0}+\frac{R^2}{2a_0^2}+\frac{R^3}{6a_0^3}+\frac{R^4}{24a_0^4}\right\}\right]^{\frac{1}{2}} = f(\varepsilon)$ attains a maximum value $f(\varepsilon) = 1$, at $R = 25a_0$, and remains unity for all $R > 25a_0$. Also, we notice that the variation in the *horizontal lift* $l = 2\pi f(\varepsilon)$, is reminiscent of elliptical curves for values $R < 25a_0$, and the horizontal lift is maximal with a value $2\pi$ for $R \geq 25a_0$, and the corresponding curves are circular in nature. Noticeably, the wave function as trivial as $\Psi_{100}$ does not exhibit any curvilinear behaviour close to our interest as discussed here.

In nut-shell, the *horizontal lift* or the *length* of a curve defined as $l = \int \left\langle \frac{\partial \Psi}{\partial t} \bigg| \frac{\partial \Psi}{\partial t} \right\rangle^{\frac{1}{2}} dt$ in the quantum geometric explorations behaves in the same manner as the function $2\pi f(\varepsilon)$ for any curve behaves in the classical geometry.

These exercises do not end here. In fact this is motivational work and we have whole lot of quantum states and quantum fields open for investigations.

## 4. Summary

We are not surprised by these exercises, as the message about the nature of the curves associated with Hydrogen like wave functions is already stored in the very eigen-functions of Hydrogen atom in terms of spherical harmonics. It is worth noticing here the exercises incorporating gravity in the Schrödinger's equation as:

$$-\frac{\hbar^2}{2m}\nabla^2\Psi + (mgx - E)\Psi = 0, \qquad (28)$$

have been found to be absolutely consistent within its bundle structure [18]. We reiterate here the famous statement by Wheeler- "Matter tells space-time how to curve, and space-time reacts back on the matter telling it how to move". We emphasize the two parts of the statement exclusively and assert that the pre-selection of the flat space-time does not guarantee that matter subjected to it will not curve it any more. The present study is an exercise to evaluate some of these effects.

The precise motivation to investigate geometric quantum mechanics for more and more classical geometric forms is to address the quest for gravity in Quantum Mechanics. Since the theory of classical gravity is basically geometric in nature and Quantum Mechanics is in no way devoid of geometry, the classical geometric forms in Quantum Mechanical explorations could be identified as signatures of gravity in Quantum Mechanics. The course of explorations in Geometric Quantum Mechanics triggered with investigations of simple geometric forms in Quantum Mechanics having traversed several milestones has indeed come a long way. And still there could be a hidden treasure of information that is yet to be retrieved and which could prove to be valuable for larger objectives such as understanding of gravity.

## Acknowledgement

The author wishes to thank Prof. A. Ashtekar, and Prof. N. Mukunda for their critical comments and suggestions. Also, the author is grateful to Institute of Mathematical Sciences (Chennai), Indian Institute of Science (Bangalore), Inter University Centre for Astronomy and Astrophysics *(IUCAA)* Pune, and *IUCAA Reference Centre* at Delhi for the facilities provided to him during his stay at these institutions.